\documentclass[preprint,nofootinbib,aps,amsfonts,superscriptaddress]{revtex4}
\usepackage{graphicx,amsmath,amssymb,bm}
\usepackage{amsfonts,amssymb,amscd,amsmath}
\usepackage{graphicx}

\newcommand\fverb{\setbox\pippobox=\hbox\bgroup\verb}
\newcommand\fverbdo{\egroup\medskip\noindent%
                        \fbox{\unhbox\pippobox}\ }
\newcommand\fverbit{\egroup\item[\fbox{\unhbox\pippobox}]}

 \newcommand\beq{\begin{equation}}
 \newcommand\eeq{\end{equation}}
 \newcommand\beqn{\begin{eqnarray}}
 \newcommand\eeqn{\end{eqnarray}}

\def\eq#1{{Eq.~(\ref{#1})}}
\def\fig#1{{Fig.~\ref{#1}}}

\begin{document}

\title{\vspace*{2cm} {\bf Comparing AdS/CFT Calculations to HERA ${\bf
      F_2}$ Data}\\[.5cm]}

\author{Yuri V.\ Kovchegov}
\affiliation{Department of Physics, The Ohio State University, Columbus, OH 43210, USA}
\author{Zhun Lu}
\affiliation{Departamento de F\'\i sica y Centro de Estudios
Subat\'omicos,\\ Universidad T\'ecnica
Federico Santa Mar\'\i a, Casilla 110-V, Valpara\'\i so, Chile}
\author{Amir H. Rezaeian}
\affiliation{Departamento de F\'\i sica y Centro de Estudios
Subat\'omicos,\\ Universidad T\'ecnica
Federico Santa Mar\'\i a, Casilla 110-V, Valpara\'\i so, Chile}
\affiliation{Institut f\"ur Theoretische Physik, Universit\"at Regensburg, D-93040 Regensburg, Germany\vspace*{1cm}}

\begin{abstract}
  We show that HERA data for the inclusive structure function
  $F_2(x,Q^2)$ at small Bjorken-$x$ and $Q^2$ can be reasonably well
  described by a color-dipole model with an AdS/CFT-inspired
  dipole-proton cross section. The model contains only three free
  parameters fitted to data. In our AdS/CFT-based parameterization the
  saturation scale varies in the range of $1 \div 3~\text{GeV}$
  becoming independent of energy/Bjorken-$x$ at very small $x$. This
  leads to the prediction of $x$-independence of the $F_2$ and
    $F_L$ structure functions at very small $x$. We provide
    predictions for $F_2$ and $F_L$ in the kinematic regions of future
    experiments. We discuss the limitations of our approach and its
  applicability region, and argue that our AdS/CFT-based model of
  non-perturbative physics could be viewed as complimentary to the
  perturbative description of data based on saturation/Color Glass
  Condensate physics.
\end{abstract}


\maketitle
                       

\date{\today}





\section{Introduction}

Experimental measurements of the proton structure function in deep
inelastic lepton-hadron scattering (DIS) at small Bjorken-$x$ have
been one of the most valuable sources of information for the
exploration of a new regime of QCD which is characterized by high
parton density. For sufficiently high energies/small Bjorken-x,
perturbative QCD predicts that gluons in a hadron wavefunction form a
Color Glass Condensate (CGC)
\cite{Gribov:1984tu,Jalilian-Marian:1997jx,Kovchegov:1999yj,Iancu:2003xm}.
The main principle of the CGC is the existence of a hard saturation
scale $Q_s$ at which nonlinear gluons recombination effects start to
become important. The saturation scale insures that the strong
coupling constant is small.

The saturation scale $Q_s$ grows rapidly with energy or a power of
$1/x$ as follows from the perturbative nonlinear small-x
Balitsky-Kovchegov (BK) \cite{Kovchegov:1999yj} and
Jalilian-Marian--Iancu--McLerran--Weigert--Leonidov--Kovner (JIMWLK)
\cite{Jalilian-Marian:1997jx} quantum evolution equations.  The BK and
JIMWLK evolution equations unitarize the linear
Balitsky-Fadin-Kuraev-Lipatov (BFKL) \cite{Kuraev:1977fs} evolution
equation at small-$x$ in the large-$N_c$ limit (BK) and beyond
(JIMWLK). In the leading logarithmic ($\ln 1/x$) approximation at
fixed coupling, the BK equation predicts that $Q_s^2(x)\sim
(1/x)^{4.6~\alpha_s}$ ($\alpha_s$ is the strong coupling)
\cite{Iancu:2002tr,Albacete:2004gw}, which is a much faster growth of
the saturation scale than one expects phenomenologically from HERA
data. On the other hand, it has been shown that next-to-leading-order
(NLO) corrections to the BFKL equation (and therefore to BK and JIMWLK
kernels) are large and negative \cite{Fadin:1998py}: they slow down
the growth of the cross sections (and, therefore, of the
  saturation scale) with energy too much for the theory to fit the
  data.  It is generally believed that the higher order corrections
to the NLO BK and JIMWLK equations should remedy this problem and
bring CGC theoretical predictions closer to the experimental data.
This idea has been recently supported by the phenomenological success
of the inclusion of running coupling corrections into the BFKL, BK and
JIMWLK equations \cite{Albacete:2007sm,rc,Albacete:2009fh}.

Another possible way to constrain higher order corrections to the
BFKL, BK and JIMWLK equations is to consider small-$x$ evolution in
the large coupling limit. At large coupling all higher order
perturbative corrections are summed up: thus the behavior of the
scattering amplitude and cross sections at strong coupling should
serve as a guide to estimate the size of higher order corrections to
the perturbative (small coupling) evolution equations. Indeed strong
coupling analytic calculations are not possible in QCD.  In light of
this, one may resort to other QCD-like theories, such as
$\mathcal{N}=4$ super Yang-Mills (YM) where one can perform
calculations in the non-perturbative limit of large `t Hooft coupling
by employing the Anti-de Sitter space/conformal field theory (AdS/CFT)
correspondence \cite{Maldacena:1997re}. Analysis of high energy
scattering amplitudes in the AdS/CFT framework was pioneered in
\cite{Janik:1999zk,Polchinski:2001tt}. Applications of AdS/CFT
techniques to DIS were further developed in \cite{Hatta:2007cs}.

Very recently, the authors of \cite{ads1} calculated the total
cross-section for a quark dipole scattering on a nucleus at high
energy for a strongly coupled $\mathcal{N} = 4$ super Yang-Mills (SYM)
theory using AdS/CFT correspondence.  The forward scattering amplitude
for the $q\bar{q}$ dipole-nucleus scattering was derived in
\cite{ads1} and exhibited an interesting feature: at high energy the
amplitude would stop growing with energy, becoming a constant. Such
phenomenon happens even for the range of dipole sizes where the
interaction is still not very strong, outside of the black disk
limit. At very small dipole sizes the amplitude continues to grow fast
with energy, in qualitative agreement with the findings of
\cite{Janik:1999zk,Polchinski:2001tt} (see \cite{ads1} for details).
The slow growth with energy of the DIS cross section found in
\cite{ads1} may allow one to identify it with the soft pomeron
contribution \cite{Donnachie:1998gm}. As such the amplitude may be
compatible to DIS data in the (presumably) non-perturbative region of
small $Q^2$. Indeed one has to keep in mind that the results of
\cite{ads1} were derived for $\mathcal{N} = 4$ SYM theory, and their
relation to QCD should be qualitative at best.

The main aim of this paper is to confront the color-dipole scattering
amplitude on a nucleus from \cite{ads1} with the available HERA data.
It is not a priori obvious whether the available data at HERA are in
the kinematics regime of validity of this model.  Given the
non-perturbative nature of the AdS/CFT approach, we expect this model
to be valid at small $x$ but also at small $Q^2$ where the
experimental data is very limited. Below we show that the HERA data
for the inclusive structure function $F_2(x,Q^2)$ for $x<6\times
10^{-5}$ and $Q^2<2.5~\text{GeV}^2$ can be well described within the
color dipole picture inspired by the AdS/CFT approach of \cite{ads1}.
We extract the saturation scale from the dipole-proton scattering
amplitude fitted to HERA data. We show that, unlike the perturbative
predictions for its behavior, the saturation scale given by the
AdS/CFT approach of \cite{ads1} becomes independent of
energy/Bjorken-x at very high energy, while being energy-dependent at
lower energies. This leads to a new phenomenon, the {\em
  $x$-independent} behavior of $F_2$ structure function at very small
$x$ and $Q^2$. We point out that qualitatively similar behavior of
$F_2$ (i.e., slowing down of the $x$-dependence at small-$x$) is
expected from the CGC approach as well \cite{Albacete:2009fh}.

The paper is organized as follows: in Sect. II we briefly recall the
color dipole description of structure function $F_2$. In Sect. III we
introduce the AdS/CFT model for the dipole-target forward scattering
amplitude. In Sect. IV we present our AdS-inspired fit to the HERA
$F_2$ data. In Sect. V we plot our fit for the $F_2$ structure
  function and extend our curves to make $F_2$ predictions for smaller
  values of $x$ than measured at HERA. We do the same for the charm
  structure function $F_2^c$. We also make predictions for the
  longitudinal structure function $F_L$ and the total photoproduction
  cross section. As a conclusion, in Sect. VI we highlight the main
results, and discuss the prospects and caveats of our model.
  

\section{Color dipole description of structure function $F_2$}

One of the most promising approaches to description of the DIS total
and diffractive lepton-proton cross sections at small $x$ has been the
color dipole factorization scheme.  In the color-dipole picture the
scattering between the virtual photon $\gamma^{\star}$ and the proton
is seen as the dissociation of $\gamma^{\star}$ into quark-antiquark
pair (the so-called $q\bar{q}$ dipole) of flavor $f$ with transverse
size $r$ which then interacts with the proton via gluon exchanges and
emissions,
\begin{equation}\label{gp}
  \sigma_{L,T}^{\gamma^*p}(Q^2,x) = \sum_f \int d^2r\,\int_0^1 dz\,
  |\Psi_{L,T}^{(f)}(r,z;Q^2)|^2\,\sigma_{q\bar{q}}(r,x),
\end{equation}
where the light-cone wavefunction $\Psi_{L,T}^{(f)}$ for
$\gamma^{\star}$ is computable in QED
\cite{Nikolaev:1990ja,gbw,Kovchegov:1999kx} with $L,T$ denoting the
longitudinal and transverse polarizations of the virtual photon:
\begin{subequations}\label{wfs}
\begin{eqnarray}
  |\Psi_{T}^{(f)}(r,z;Q^2)|^2 \, & = & \,
  \frac{\alpha_{EM} \, N_c}{2\, \pi^2} \, \sum\limits_f \, e_f^2 \, 
  \left\{ a_f^2 \, [K_1 (r \, a_f)]^2 \, [z^2 + (1 - z)^2] 
  + m_f^2 \, [K_0 (r \, a_f)]^2 \right\}, \ \label{wav}\\
  |\Psi_{L}^{(f)}(r,z;Q^2)|^2 \, & = & \, 
  \frac{\alpha_{EM} \, N_c}{2 \, \pi^2} \, \sum\limits_f e_f^2 \, 
  \left\{ 4 \, Q^2 z^2 (1 - z)^2 \, [K_0 (r \, a_f)]^2 \right\}. \label{wavL}
\end{eqnarray} 
\end{subequations}
Here $z$ is the fraction of the light cone momentum of the virtual
photon carried by the quark, $m_f$ is quark mass, $a_f^2 = z \, (1-z)
\, Q^2 + m_f^2$, $\alpha_{EM}$ is the electromagnetic coupling
constant, $e_f$ is the electric charge of a quark with flavor $f$, and
$N_c$ denotes the number of colors. Below, we will first follow
  \cite{gbw} and use three light quark flavors only with $m_u = m_d =
  m_s =140$~MeV. Then, we will also consider a case with three light
  flavors and a charm quark with mass $m_c=1.4$~GeV. To estimate the
  effect of light quark masses, we will also consider a case with
  massless light quarks. For the light quarks, the gluon density is
  evaluated at $x = x_{Bj}$ (Bjorken-x), while for charm quarks we
  take $x = x_{Bj} \, (1 + 4 m^2_c/Q^2)$.

The $q\bar{q}$ dipole-proton cross-section $\sigma_{q\bar{q}}(r,x)$
incorporates QCD effects. It is usually written as an integral of the
imaginary part of the forward scattering amplitude $N ({\bf r}, {\bf
  b}, x)$ over the impact parameter $\bf b$ \cite{Kovchegov:1999yj}:
\begin{equation}
  \label{sigmaN}
  \sigma_{q\bar{q}} (r,x) \, = \, 2 \, \int d^2 b \, N ({\bf r}, {\bf
  b}, x), 
\end{equation}
where bold letters denote two-dimensional vectors in transverse plane.
Following the usual approach we will neglect the $b$-dependence in $N$
making the integral in \eq{sigmaN} trivial giving the proton's
transverse area factor: $\sigma_{q\bar{q}}(r,x) \equiv \sigma_0 \, N
(r,x)$.

The proton structure function $F_2$ and the longitudinal structure function $F_L$ can be written in terms of
$\gamma^{\star}p$ cross-section,
\begin{eqnarray}
F_2(Q^2,x) &=& \frac{Q^2}{4\pi^2\alpha_{EM}} 
\left[\sigma_L^{\gamma^*p}(Q^2,x)+\sigma_T^{\gamma^*p}(Q^2,x)\right],\label{f2}\\
F_L(Q^2,x) &=& \frac{Q^2}{4\pi^2\alpha_{EM}}\sigma_L^{\gamma^*p}(Q^2,x). \ \label{FL}
\end{eqnarray}
The contribution of the charm quark to the wave functions in Eqs.
(\ref{wfs}) feeds into Eqs.~(\ref{gp}) and (\ref{f2}) directly giving
the charm structure function $F_2^c$. In the CGC framework the
dipole-proton forward scattering amplitude $N$ can be found by solving
BK or JIMWLK evolution equations
\cite{Lublinsky:2001yi,Albacete:2009fh}.  Alternatively there exist
many different phenomenological approaches to model both CGC and
non-perturbative effects in the dipole cross-section or amplitude
which can be then tested against the HERA data, see \cite{gbw-g} and
references therein.  Here, we show that the AdS/CFT-inspired
color-dipole model of \cite{ads1} predicts a new scaling behavior for
the proton structure function at very small $x$ and $Q^2$ in a region
where there is no experimental data yet and argue that future
experimental measurement of $F_2$ in this region can be used to test
the model.


\section{AdS/CFT color dipole model}

The forward scattering amplitude $N$ of a $q\bar{q}$ dipole on a large
nuclear target (with atomic number $A$) at high-energy for a strongly
coupled $\mathcal{N}=4$ super Yang-Mills theory employing AdS/CFT
correspondence was derived in \cite{ads1} and has the following form:
\begin{eqnarray} 
  N(r,s)&=&1-\exp\Big[-\frac{a_0}{s}\left(\frac{c_0^2 r^2 }{\rho^3}+\frac{2}{\rho}-2\sqrt{s}\right)\Big], \label{fas0}\\
\rho&=&c_0 \, r \, \sqrt{\frac{1}{3m\Delta} +  \Delta}, \\
\Delta&=& \Big[\frac{1}{2m}-\sqrt{\frac{1}{4m^2}-\frac{1}{27m^3}}\Big]^{1/3}, \label{del}\\
m&=&c_0^4 \, r^4 \, s^2. \label{mmm}
\end{eqnarray}
The parameter $c_0$ in the above equations is a constant which relates
the transverse dipole size $r$, the collision energy $\sqrt{s}$ and
the maximum extent of the string in the $z$-direction labeled by
$z_{max}$ \cite{ads1},
\begin{equation}
c_0 r=z_{max}\sqrt{1-s^2 z^4_{max}}, \label{c01}
\end{equation}
where the value of $c_0$ is given by\footnote{From Eq.~(\ref{c01}) one
  can immediately recover the case considered by Maldacena
  \cite{Maldacena:1998im} for the shape of a static Wilson loop in an
  empty AdS$_5$ space by putting $s=0$.}
\begin{equation}
c_0=\frac{\Gamma^2(\frac{1}{4})}{(2\pi)^{3/2}}. \label{c0}
\end{equation}
The parameter $a_0$ in Eq.~(\ref{fas0}) is given by
\begin{equation}
a_0=\frac{\sqrt{\lambda_{YM}}A^{1/3}\Lambda}{\pi c_0\sqrt{2}},\label{a0}
\end{equation}
where $\lambda_{YM}=g^2_{\text{YM}} N_c$ denotes the `t Hooft coupling
with $g_{\text{YM}}$ the Yang-Mills coupling constant. The parameter
$\Lambda$ can be identified as the transverse momentum scale
\cite{ads1}.  Note that in Eq.~(\ref{del}) $\Delta$ can be imaginary
for small $m$, but the parameter $\rho$ is always real.

One can also rewrite the dipole amplitude Eq.~(\ref{fas0}) as a
function of Bjorken-$x$.  To simplify and approximate the $r$-integral
in \eq{gp} we relate the virtuality of the photon $Q$ to the dipole
size $Q= b_0/r$ where the parameter $b_0$ will be determined from a
fit to the data.  Therefore, the Bjorken-$x$ variable in DIS
becomes\footnote{Note that we ignore the proton mass in the Bjorken-x
  definition since its effects in the kinematic region of our interest
  is negligible and will not change the results.}
\begin{equation}
x\equiv\frac{Q^2}{s+Q^2}\equiv \frac{b_0^2}{b_0^2+s \, r^2}. \label{as1}
\end{equation}
By using the above relation, one can rewrite the $q\bar{q}$
dipole-nucleus amplitude defined in Eq.~(\ref{fas0}) as a function of
$x$ and $r$,
\begin{eqnarray}
  N(r,x)& =&1-\exp\Big[-\frac{\mathcal{A}_0 \, x \, r}{\mathcal{M}_0^2(1-x)\pi \sqrt{2}}\left(\frac{1}{\rho_m^3}+\frac{2}{\rho_m}-2\mathcal{M}_0\sqrt{\frac{1-x}{x}}\right)\Big], \label{fas1}\
\end{eqnarray}
with
\begin{eqnarray}
\rho_m&=&  \begin{cases}
 (\frac{1}{3m})^{1/4}\sqrt{2\cos(\frac{\theta}{3})}  & : m \le \frac{4}{27} \nonumber\\
     \sqrt{\frac{1}{3m\Delta} +  \Delta} & :m> \frac{4}{27}
  \end{cases},  \nonumber\\
\Delta&=& \Big[\frac{1}{2m}-\sqrt{\frac{1}{4m^2}-\frac{1}{27m^3}}\Big]^{1/3} \nonumber\\
m&=&\frac{\mathcal{M}_0^4(1-x)^2}{x^2}, \nonumber\\
\cos(\theta)&=&\sqrt{\frac{27m}{4}}, \label{not}\
\end{eqnarray}
where we defined $\mathcal{M}_0=b_0c_0$ and
$\mathcal{A}_0=\sqrt{\lambda_{YM}} \, \Lambda$. The impact-parameter
integrated $q\bar{q}$ dipole cross-section on a proton target is then
related to the dipole amplitude via $\sigma_{q\bar{q}}(r,x)=\sigma_0
\, N(r,x)$.

As a comparison to other dipole models, we will cross-check our
results with the popular GBW color dipole model proposed by
Golec-Biernat and W\"usthoff \cite{gbw}. This model is able to
describe DIS data with the dipole cross-section parametrized as
\begin{equation}
\sigma_{q\bar{q}}^{\text{GBW}}(x,\vec{r}) \, = \, \sigma_{0} \, 
\left(1-e^{-r^{2}Q_{s}^{2}(x)/4}\right), \label{gbw}
\end{equation}
where $x$-dependence of the saturation scale is given by
\begin{equation}
Q_s^{\text{GBW}}(x)\equiv Q_s(x)= \, 
\left(\frac{x_0}{x} \right)^{\lambda/2}~\text{GeV}. \label{sgbw}
\end{equation}
We have not assumed anything about the functional form of the
saturation scale in the dipole amplitude (\ref{fas1}).  Note that
there is no unique definition for the saturation scale in literature.
Following Refs.~\cite{gbw,di1,di2,di3} we define a saturation scale
$Q_s^2=2/r_s^2$ as a momentum scale at which the $q\bar{q}$ dipole
scattering amplitude $N$ becomes sizable
\begin{equation}
N(r_s=\sqrt{2}/Q_s,x)=\mathcal{N}_0 \equiv 1 - e^{-1/2} \approx 0.4.  \label{defs}
\end{equation}
For the GBW model, this definition coincides with the saturation scale
$Q_s$ defined in Eq.~(\ref{sgbw}).  Similarly, the saturation scale in
AdS/CFT dipole model (\ref{fas1}) is then defined as
\begin{equation} 
Q_s^{\text{AdS}}(x)=\frac{2 \, \mathcal{A}_0 \, x}{\mathcal{M}_0^2 \, (1-x) \, \pi} \, 
\left(\frac{1}{\rho_m^3}+\frac{2}{\rho_m}-2\mathcal{M}_0\sqrt{\frac{1-x}{x}}\right). \label{qads}
\end{equation}
Note that the AdS/CFT dipole scattering amplitude $N$ from
Eq.~(\ref{fas1}) with the saturation scale from \eq{qads} exhibits the
property of geometric scaling \cite{Stasto:2000er}: it is a function
of $r \, Q_s^{\text{AdS}}(x)$ only, $N (r,x) = 1- \exp [- {r \,
  Q_s^{\text{AdS}}(x) /(2 \sqrt{2})}]$.  Moreover, the anomalous
dimension in the AdS/CFT dipole model is $\gamma_s=0.5$ which is
rather close to the value of $0.44$ obtained from the numerical
solution of the BK equation \cite{bk-g}. Thus in many ways our AdS/CFT
inspired model is similar to the predictions of CGC. The main
difference is in the $x$-dependence of the saturation scale
$Q_s^{\text{AdS}}(x)$, which we will discuss shortly.



\section{Fit to HERA $F_2$ Data}

In this section, we confront the AdS/CFT color-dipole with the
experimental data from DIS and test its validity by investigating
whether its free parameters can be fitted to the experimental
measurements of the proton structure function $F_2$.

In the dipole amplitude given by Eqs.~(\ref{fas1}) we take
$\mathcal{M}_0=b_0 c_0$ to be a free parameter since the value of
$b_0$ is not known. The parameters $b_0$ and $c_0$ always appear only
as a product denoted by $\mathcal{M}_0$ and cannot be taken in the
fitting as two independent parameters. By taking $\mathcal{M}_0$ as a
free parameter, we also allow the parameter $c_0$ to deviate from its
value obtained from the AdS/CFT approach. This is motivated by the
fact that the value of $c_0$ given by \eq{c0} is true for
$\mathcal{N}=4$ SYM theory, and is likely to be different for QCD.
The parameter $\mathcal{A}_0=\sqrt{\lambda_{YM}} \, \Lambda$ in the
AdS/CFT dipole model appears as an overall factor in the saturation
scale in Eq.~(\ref{qads}): it can be taken as another free parameter
in the fit.  As $\lambda_{YM}$ and $\Lambda$ only appear together in
$\mathcal{A}_0$ we put $\Lambda=1$~GeV throughout this paper and vary
$\lambda_{YM}$. We examine different cases with $\lambda_{YM}=5, 10,
20,30$ and $40$. The other two free parameters $\mathcal{M}_0$ and
$\sigma_0$ in the AdS/CFT dipole model will be determined from a fit
to the DIS data.  Notice that the GBW dipole model given by
Eq.~(\ref{gbw}) also has $3$ unknown parameters: $x_0, \lambda$, and
$\sigma_0$.

\begin{table}
  \begin{tabular}{cccc|c} 
 \hline\hline AdS/CFT dipole model& $\lambda_{YM}$ & $\mathcal{M}_0/ 10^{-3}$ &  $\sigma_0[\text{mb}]$ & $\chi^2/\text{d.o.f.} $ \\
  \hline 
   $x \in[6.2\times 10^{-7},10^{-4}]$ &$5$ & $9.85 $ & $31.164$ & $110.70/78 =1.42  $ \\ 
  $x \in[6.2\times 10^{-7},10^{-4}]$ &$20$ & $ 6.36$ & $ 22.65$ & $ 141.12/78 = 1.81 $ \\ 
\hline
  $x \in[6.2\times 10^{-7},6\times 10^{-5}]$&$5$ &    $10.114$& $30.97$ & $44.24/60 = 0.74$ \\ 
$x \in[6.2\times 10^{-7},6\times 10^{-5}]$&$10$ &    $8.16$& $26.08$ & $49.22/60 = 0.82$ \\ 
  $x \in[6.2\times 10^{-7},6\times 10^{-5}]$&$20$ &    $6.54$& $22.47$ & $ 55.195/60 = 0.92$ \\ 
  $x \in[6.2\times 10^{-7},6\times 10^{-5}]$&$30$ &    $5.72  $& $ 20.80 $ & $58.87/60 = 0.98  $ \\ 
  $x \in[6.2\times 10^{-7},6\times 10^{-5}]$&$40$ &    $5.20 $& $ 19.78$ & $ 61.47/60 = 1.024$ \\ 
   \hline\hline
  \end{tabular} 
     \caption{Parameters of the AdS/CFT dipole model from Eq.~(\ref{fas1})
  determined from a fit to $F_2$ data reported by ZEUS in two Bjorken $x$
  bins. The value of quark mass $m_{u,d,s}=140$ MeV is taken in all the fits. 
(Here we consider only three light flavors.) The data for the first two rows and the 
rest are within $Q^2/\text{GeV}^2\in [0.045,6.5]$ and $Q^2/\text{GeV}^2\in [0.045,2.5]$ 
respectively. 
   }
\end{table}

\begin{table}
 \begin{tabular}{cccc|c} 
 \hline\hline  GBW dipole model&
  $x_0/10^{-4}$ & $\lambda$ & $\sigma_0[\text{mb}]$ &$\chi^2/\text{d.o.f.} $ \\ \hline 
   $x \in[6.2\times 10^{-7},10^{-4}]$ & $ 2.225$ & $0.299 $& $22.77$ & $ 63.09/78 =0.81
  $ \\ 
\hline
  $x \in[6.2\times 10^{-7},6\times 10^{-5}]$ & $2.371$ & $0.368$ &
  $21.13$ & $39.35/60 = 0.66  $ \\ \hline\hline 
   \end{tabular}
  \caption{Parameters of the GBW color dipole model
  determined from a fit to $F_2$ data from ZEUS in two Bjorken $x$
  bins. The value of quark mass $m_{u,d,s}=140$ MeV is taken for both fits. 
The data for the first and the second row are within $Q^2/\text{GeV}^2\in [0.045,6.5]$ 
and $Q^2/\text{GeV}^2\in [0.045,2.5]$ respectively. 
   }
\end{table}

We shall use HERA data from ZEUS \cite{zu1,zu2,zu3,zu4} measurement of
$F_2$. Following the earlier analysis
Refs.~\cite{di1,di2,di3,di4,di5}, we do not include the H1 data in
order to avoid introducing extra normalization parameters relating
ZEUS and H1 data. The AdS/CFT color dipole model is motivated by
non-perturbative QCD and could only be applicable at small $Q^2$.
Therefore, we are interested in small $x$ and $Q^2$ where most data is
from ZEUS. Unfortunately the experimental data points for the
structure function at very small $x$ and $Q^2$ are very limited.

\begin{table}
  \begin{tabular}{ccccc|c} 
 \hline\hline $m_c[\text{GeV}]$ & $m_{u,d,s}[\text{MeV}]$& $\lambda_{YM}$ & $\mathcal{M}_0/ 10^{-3}$ &  $\sigma_0[\text{mb}]$ & $\chi^2/\text{d.o.f.} $ \\
  \hline 
 $-$& $140$&  $\bf{10}$ &    $\bf{8.16}$& $\bf{26.08}$ & $\bf{ 49.22/60 = 0.82}$ \\ 
  $-$& $140$    &$\bf{20}$ &    $\bf{ 6.54}$& $\bf{22.47}$ &$ \bf{55.20/60 = 0.92}$ \\ 

  $-$& $0$&  $\bf{10}$ &    $\bf{ 10.81}$& $\bf{21.92}$ & $\bf{ 36.77/60 = 0.61}$ \\ 
  $-$& $0$    &$\bf{20}$ &    $\bf{8.14 }$& $\bf{19.29 }$ &$ \bf{ 37.84/60 = 0.63}$ \\

  $1.4$& $140$&  $\bf{10}$ &    $\bf{7.66}$& $\bf{24.72}$ & $\bf{61.66/60=1.03 }$ \\ 
  $1.4$& $140$    &$\bf{20}$ &    $\bf{6.16}$& $\bf{21.31}$ &$ \bf{70.99/60=1.18}$ \\ 

  $1.4$& $0$&  $\bf{10}$ &    $\bf{9.84}$& $\bf{20.79}$ & $\bf{39.10/60 =0.65 }$ \\ 
  $1.4$& $0$    &$\bf{20}$ &    $\bf{7.51}$& $\bf{18.29}$ &$ \bf{45.07/60=0.75 }$ \\ 

   \hline\hline \end{tabular} \caption{Parameters of the AdS/CFT
   dipole model from Eq.~(\ref{fas1}) determined from a fit to $F_2$
   data reported by ZEUS. We now also include charm quarks: the value of quark
   masses used in the fits are given in the table. The data used are
   within $x \in[6.2\times 10^{-7},6\times 10^{-5}]$ and
   $Q^2/\text{GeV}^2\in [0.045,2.5]$. }
\end{table}

Note that the GBW model is motivated by the perturbative QCD and its
validity at very small $Q^2$ is questionable, though it can be
extended to higher $Q^2$ if the full DGLAP evolution is used. For the
same reason, in the earlier analysis of the GBW model, the data below
$Q^2=0.25~\text{GeV}^2$ was not included in the fitting \cite{gbw},
although inclusion of those data does not have a significant effect on
the parameters obtained from the fit (see also table II). Here, we use
the GBW model only as a benchmark in order to compare our results with
a perturbatively motivated dipole model.

\begin{table}
  \begin{tabular}{ccccc|c} 
 \hline\hline $m_c[\text{GeV}]$ & $m_{u.d,s}[\text{MeV}]$& $\lambda_{YM}$ & $\mathcal{M}_0/ 10^{-3}$ &  $\sigma_0[\text{mb}]$ & $\chi^2/\text{d.o.f.} $ \\
  \hline 
  $-$& 140 &$20$ & $ 6.36$ & $ 22.65$ & $ 141.12/78 = 1.81 $ \\ 
 $-$&  0 &$20$ & $7.65 $ & $ 19.53 $ & $99.79/78 = 1.28 $ \\ 
\hline

   \hline\hline
  \end{tabular} 
     \caption{The same fit parameters as in Table III, but for 
ZEUS data taken in a slightly broader $x$-range, 
$x \in[6.2\times 10^{-7},10^{-4}]$ and $Q^2/\text{GeV}^2\in [0.045,6.5]$.
   }
\end{table}

The resulting parameters of the AdS/CFT and the GBW color-dipole
models and $\chi^2$ values obtained from the fit in which we
  consider only three light flavors with the quark mass $m_f=140$ are
presented in tables I and II for the same data bin. From table II, it
is seen that the parameters of the GBW model obtained from the fit to
the data bin of $Q^2/\text{GeV}^2\in [0.045,6.5]$ are very similar to
those in the case when one takes all the data within
$Q^2/\text{GeV}^2\in [0.25,45]$ \cite{di2}. However, the value of the
intercept $\lambda$ increases from $\lambda=0.299$ to $\lambda=0.368$
when we limit the data to a lower virtuality $Q^2/\text{GeV}^2\in
[0.045,2.5]$. The value of $\lambda\approx 0.25-0.30$ is consistent
with perturbative predictions based on small-$x$ evolution with
running coupling and other higher order corrections
\cite{lqcd,Albacete:2009fh,Albacete:2007sm,Iancu:2002tr,lbk}.  The
quality of the fit based on the AdS/CFT color-dipole model is very
sensitive to the upper bound of the given Bjorken-$x$ bin. This can be
seen in table I where including data with $x \approx 10^{-4}$
dramatically increases $\chi^2$ and worsens the fit.  This is in
contrast to the GBW model which gives a surprisingly good fit for a
wide range of $x$ (see also table II). We show in table I that a good
fit with (with $\chi^2<1$) for the AdS/CFT dipole amplitude can be
found for the current available data within $x \in[6.2\times 10^{-7},6
\times 10^{-5}]$ and $Q^2/\text{GeV}^2\in [0.045,2.5]$. Note that
currently there is no experimental data below the lower $x$ and $Q^2$
bound we have taken, and also there is no experimental data for $F_2$
at large $Q^2$ but very small $x$. In table I, we also show the
results of the fit to the same data bin for different values of
$\lambda_{YM}$.  Notice that our model is valid in the
non-perturbative regime, therefore we do not expect a very small value
of $\lambda_{YM}$. However, with a smaller $\lambda_{YM}$ we can relax
the upper bound on the $x$-bin and find a good fit for even larger
$x$. It is also seen from table I that for a wide range of
$\lambda_{YM} \ge 20$, the results of the fit change only little.

\begin{table}
  \begin{tabular}{ccc|c} 
 \hline\hline $\lambda_{YM}$ & $c_0$ &  $\sigma_0[\text{mb}]$ & $\chi^2/\text{d.o.f.} $ \\
  \hline 
          $5$ & $0.00583$ & $40.55 $ & $ 62.61/60 = 1.04 $ \\ 
          $10$ & $0.00440$ & $ 36.30$ & $77.17/60 = 1.29 $ \\ 
          $20$ & $0.00324$ & $33.58 $ & $ 92.11/60 = 1.53 $ \\ 
  \hline\hline
  \end{tabular} 
     \caption{Parameters of the $s$-dependent AdS/CFT dipole model from Eq.~(\ref{fas0})
  determined from a fit to $F_2$ data from ZEUS. Here we restrict the analysis to the 
three light flavors: the value of light quark mass $m_{u,d,s}=140$ MeV is taken in all 
three fits. The data are within $x \in[6.2\times 10^{-7},6\times 10^{-5}]$ and 
$Q^2/\text{GeV}^2\in [0.045,2.5]$. 
   }
\end{table}

While the smaller values of $\lambda_{YM}$ appear to give better
description of the $F_2$ data using our AdS/CFT ansatz, one has to
keep in mind that AdS/CFT correspondence is valid for $\lambda_{YM}
\gg 1$. We therefore can not use very small $\lambda_{YM}$ in the fit,
as the whole underlying theoretical approach of \cite{ads1} would
reach its limit of applicability. At smaller $\lambda_{YM}$ higher
order string excitations become important introducing $o\left(
  1/\sqrt{\lambda_{YM}} \right)$ corrections to the single-pomeron
intercept \cite{Polchinski:2001tt} and probably to the rest of the
expression (\ref{fas0}). To (roughly) quantify how small the coupling
$\lambda_{YM}$ can be with our ansatz (\ref{fas0}) still remaining
dominant we notice that string excitations corrections calculated in
the second reference in \cite{Polchinski:2001tt} modify the pomeron
intercept in the amplitude from $2$ to $2 - (2/\sqrt{\lambda_{YM}})$.
For the correction to be small one needs $\sqrt{\lambda_{YM}} \gg 1$.
In our analysis here we therefore restrict $\lambda_{YM}$ to be
$\lambda_{YM} \ge 5$. Indeed in QCD at low-$Q^2$ one has $\lambda_{YM}
= g_{YM}^2 \, N_c \approx 2^2 \, \times \, 3 = 12$, which is in the
range of $\lambda_{YM}$ considered in our fits.  Values of
$\lambda_{YM}$ for QCD as low as $5.5$ in AdS/CFT framework have been
considered in the literature \cite{Gubser:2006qh} to describe RHIC
heavy ion data.

In table III, we present our fits for the AdS/CFT dipole model in
  the presence of charm quark. It can be seen that while the inclusion
  of charm quark slightly worsens the fit, nevertheless $\chi^2$ is
  still in the acceptable range. In tables III, IV to investigate the
importance of the value of light quark masses, we also show the
results of the fit (to two different data bins for the two
  tables) in which the light quark masses are taken to be zero,
$m_{u,d,s}=0$. For comparison, we also show the results of the fit
obtained with $m_{u,d,s}=140$~MeV for the same data bins.  It can be
seen that taking $m_{u,d,s}=0$ improves the fit somewhat. One has to
keep in mind that the AdS/CFT dipole amplitude (\ref{fas0}) was
calculated in \cite{ads1} for very heavy dressed (constituent) quarks:
extrapolating it to massless quarks pushes \eq{fas0} closer to the
theoretical limit of its applicability. In practice, due to the
largeness of the saturation scale in this model (see below), the
structure function $F_2$ is not very sensitive to light quark masses
in the range of $m_{u,d,s}$ that we consider. One may expect
  that, since the AdS/CFT calculation of \cite{ads1} was done for
  heavy quarks, the fit should improve with the inclusion of charm.
  However, note that large charm quark mass makes QCD coupling small
  making corresponding QCD physics more perturbative. As the AdS/CFT
  calculation we are using is valid for large coupling only, inclusion
  of charm also pushes the model to the limit of its applicability.
  This could be the reason the fit gets slightly worse when we include
  charm quark mass. Another potential danger of including heavy flavor
  is in the fact that they shorten the typical coherence length of the
  quark dipole, potentially making it smaller than the size of the
  proton and invalidating the dipole approach altogether. This is a
  problem common to all dipole models.

The value of $\mathcal{M}_0=b_0 c_0$ obtained from the fit (in tables
I, III and IV) is surprisingly small. The parameter $b_0$ relates the
virtuality of the photon to the dipole size, and one expects it to be
of order of one. On the other hand, the parameter $c_0$ obtained from
the AdS/CFT approach Eq.~(\ref{c0}) is $0.83$ in $\mathcal{N}=4$ SYM
theory. In order to clarify whether the smallness of $\mathcal{M}_0$
should be associated with $c_0$ or with $b_0$ we employed
$s$-dependent AdS/CFT dipole model defined in
Eqs.~(\ref{fas0}-\ref{mmm}) which has the parameter $c_0$ but no
$b_0$. (That is we undid the numerical simplification we had made by
writing $Q=b_0/r$ to define Bjorken $x$ in \eq{as1}.) We first tried
to keep the parameter $c_0$ fixed as given in the AdS/CFT approach.
However, this did not lead to a good fit for a wide range of
$\lambda_{YM}$.  Then, arguing that $c_0$ should be different in QCD
as compared to $0.83$ in $\mathcal{N}=4$ SYM theory, we considered the
parameter $c_0$ to be a free parameter and determined it from the fit.
In table V, we show the results of the fit for different fixed values
of $\lambda_{YM}$. It is seen that generally the preferred value of
$c_0$ from the fit is two orders of magnitude smaller than the AdS/CFT
value.  This is consistent with the smallness of $\mathcal{M}_0$
obtained from the $x$-dependent AdS/CFT dipole model.  Therefore, the
smallness of $\mathcal{M}_0$ is due to the smallness of the parameter
$c_0$ preferred by HERA data.

\begin{figure}[!t]
  \includegraphics[width=8 cm] {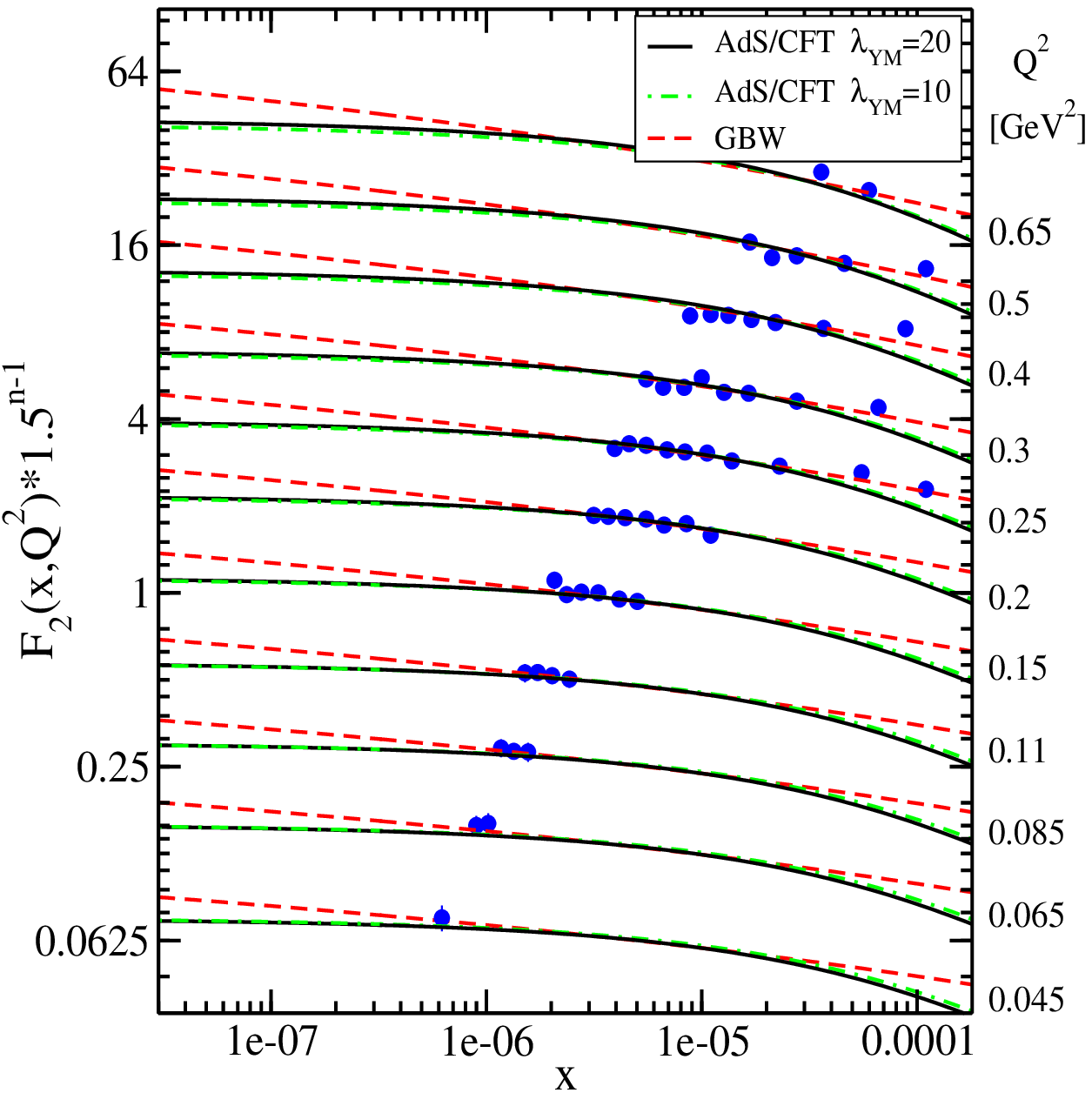} \includegraphics[width=8
  cm] {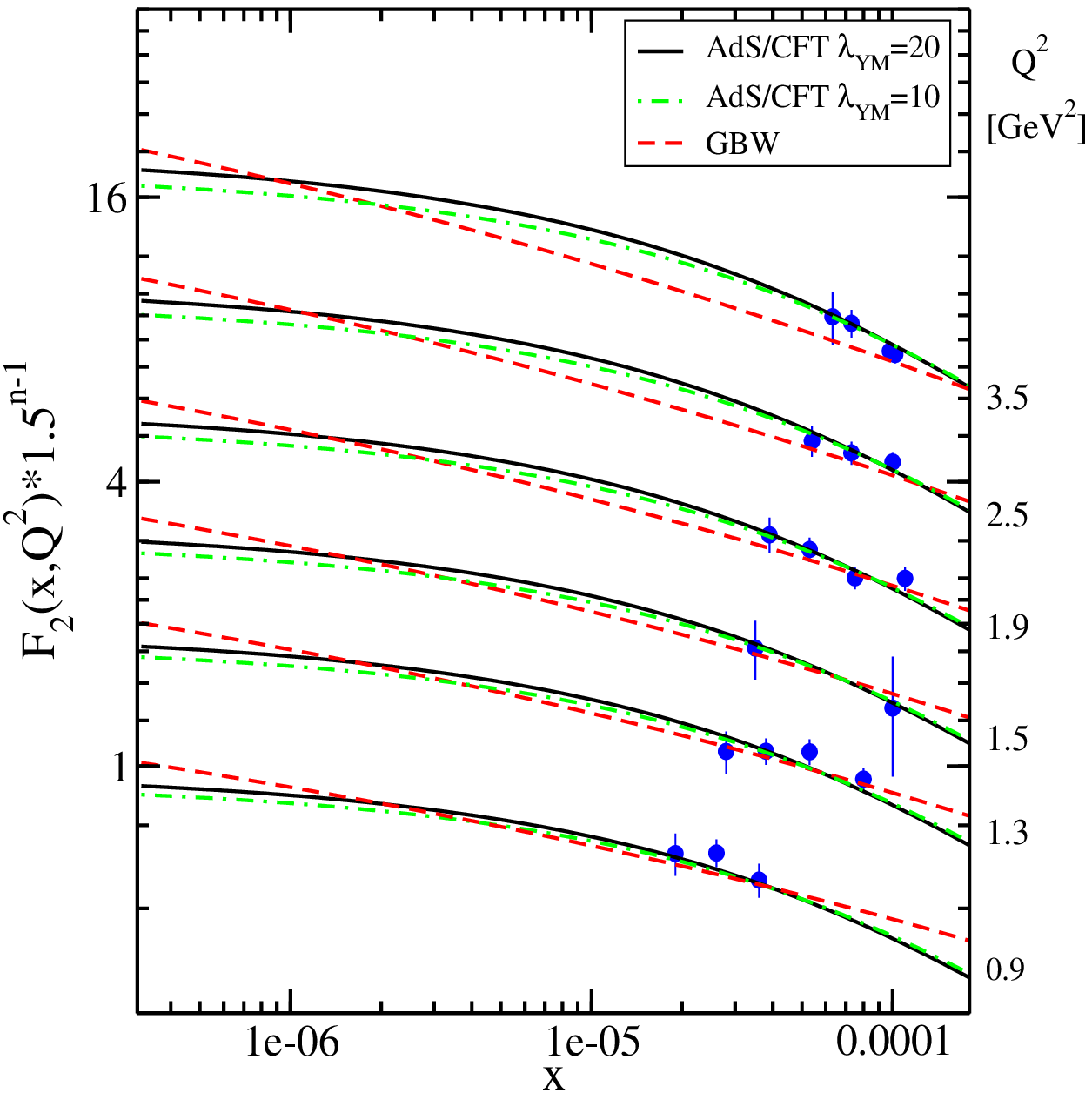}
 \caption{Results of our AdS/CFT-based fit to the
   proton structure function $F_2$. We used the fits to the data within $Q^2/\text{GeV}^2\in [0.045,2.5]$ given in tables I
   (with $\lambda_{YM}=10$ and $20$) and II (second row) for the AdS/CFT and GBW
   dipole models, respectively. The $Q^2[\text{GeV}^2]$ value
   corresponding to each curve is shown on the right margin of each
   panel. For clarity, the successive curves in both panels have
   been scaled by powers of $1.5$ from bottom to top $(n=1,2,3 \ldots
   )$.}
   \label{fig1}
\end{figure}

\section{Plots and predictions}

\begin{figure}[!t]
  \includegraphics[width=8 cm] {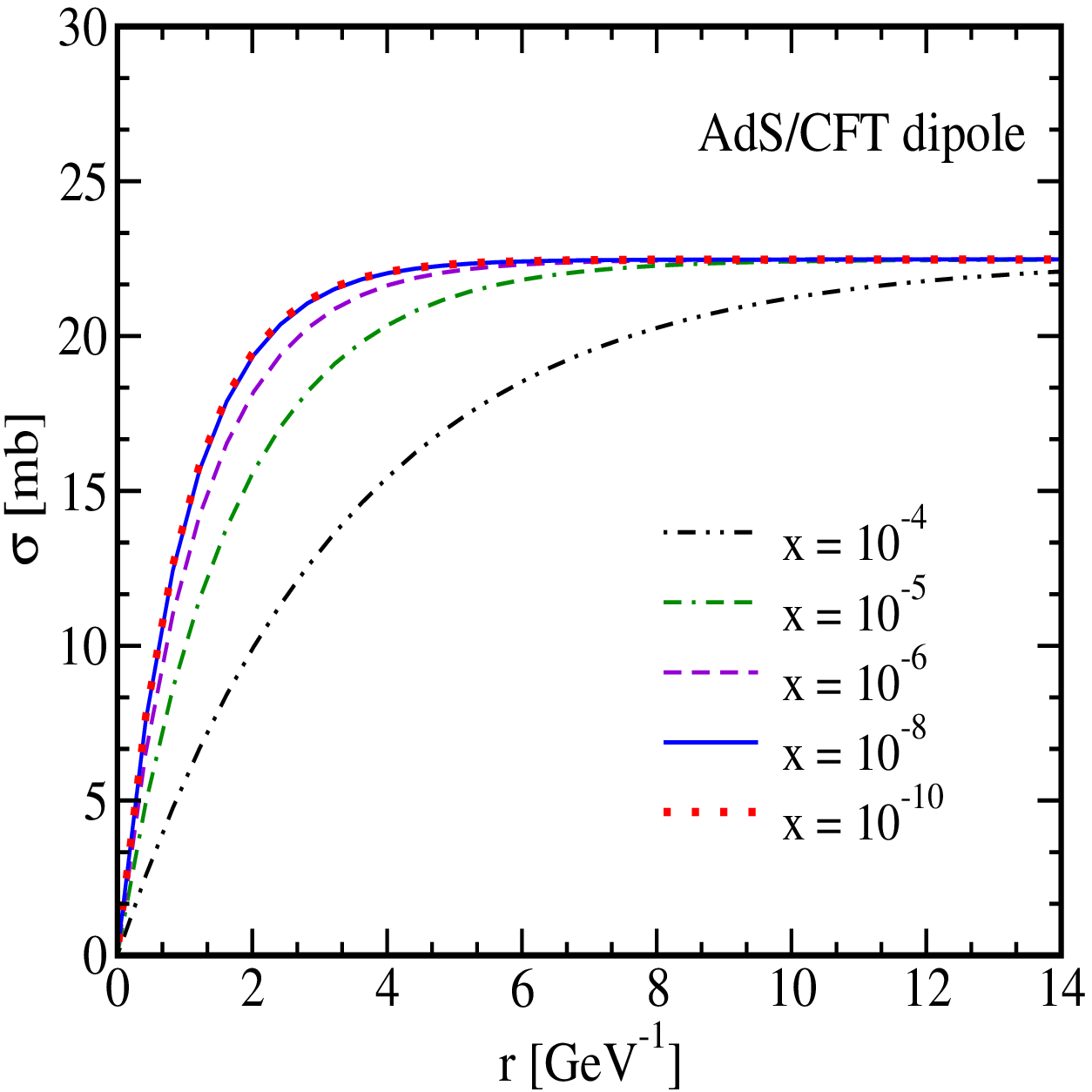}
  \includegraphics[width=8 cm] {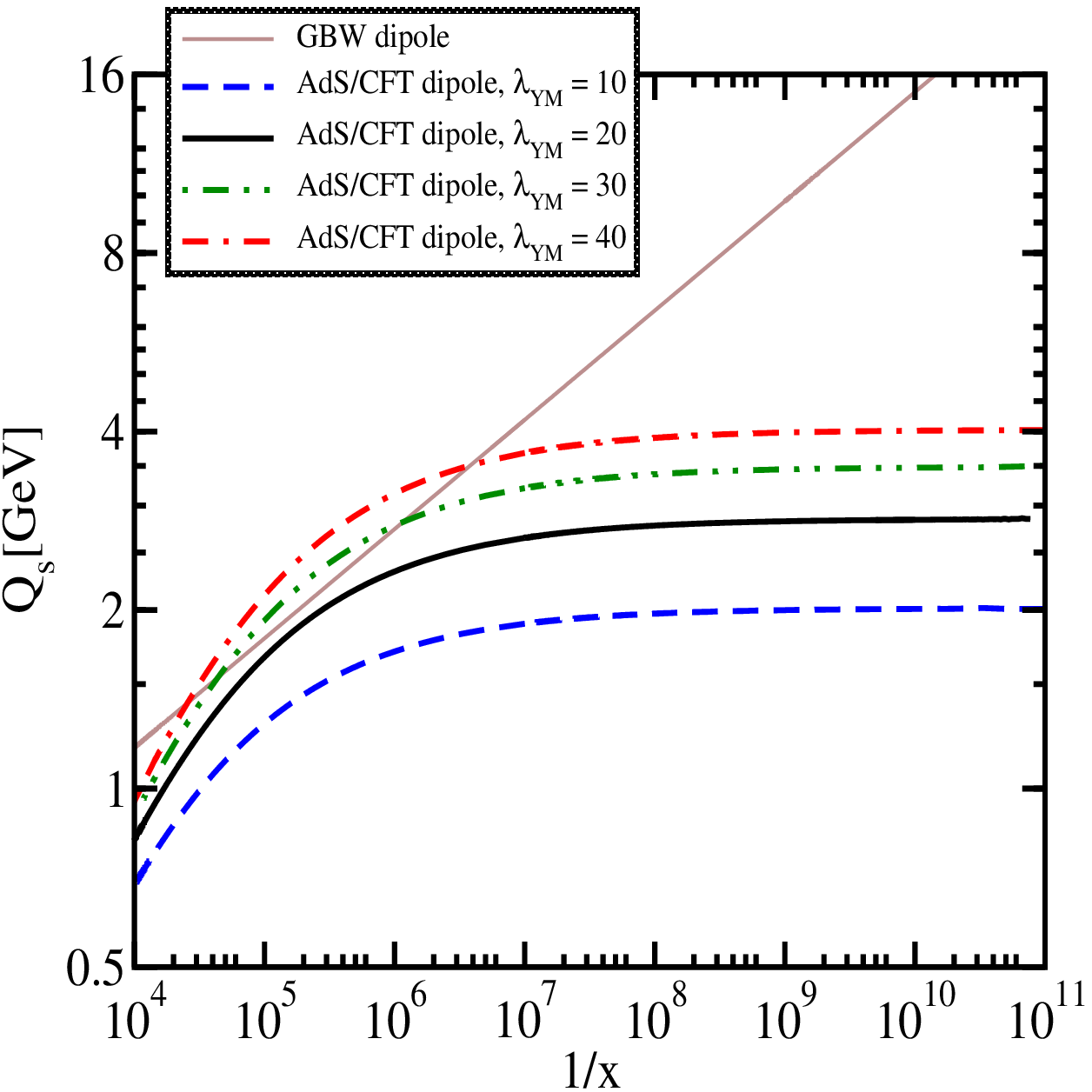} 
   \caption{ Left panel: the
     AdS/CFT dipole cross-section obtained from the fit given in table
     III for $\lambda_{YM}=20$ and $m_q=140$ MeV at various fixed
     Bjorken-$x$ as a function of the dipole size $r$. Right panel:
     the AdS/CFT and the GBW saturation scales $Q_s(x)$ [GeV] as
     functions of x. We used the fits given in tables III (with
     $m_{u,d,s}=140$~MeV and without charm quark) for the AdS/CFT
     model and in table II for the GBW dipole model.
     \label{fig2}}
\end{figure}

\begin{figure}[!t]
  \includegraphics[width=8 cm] {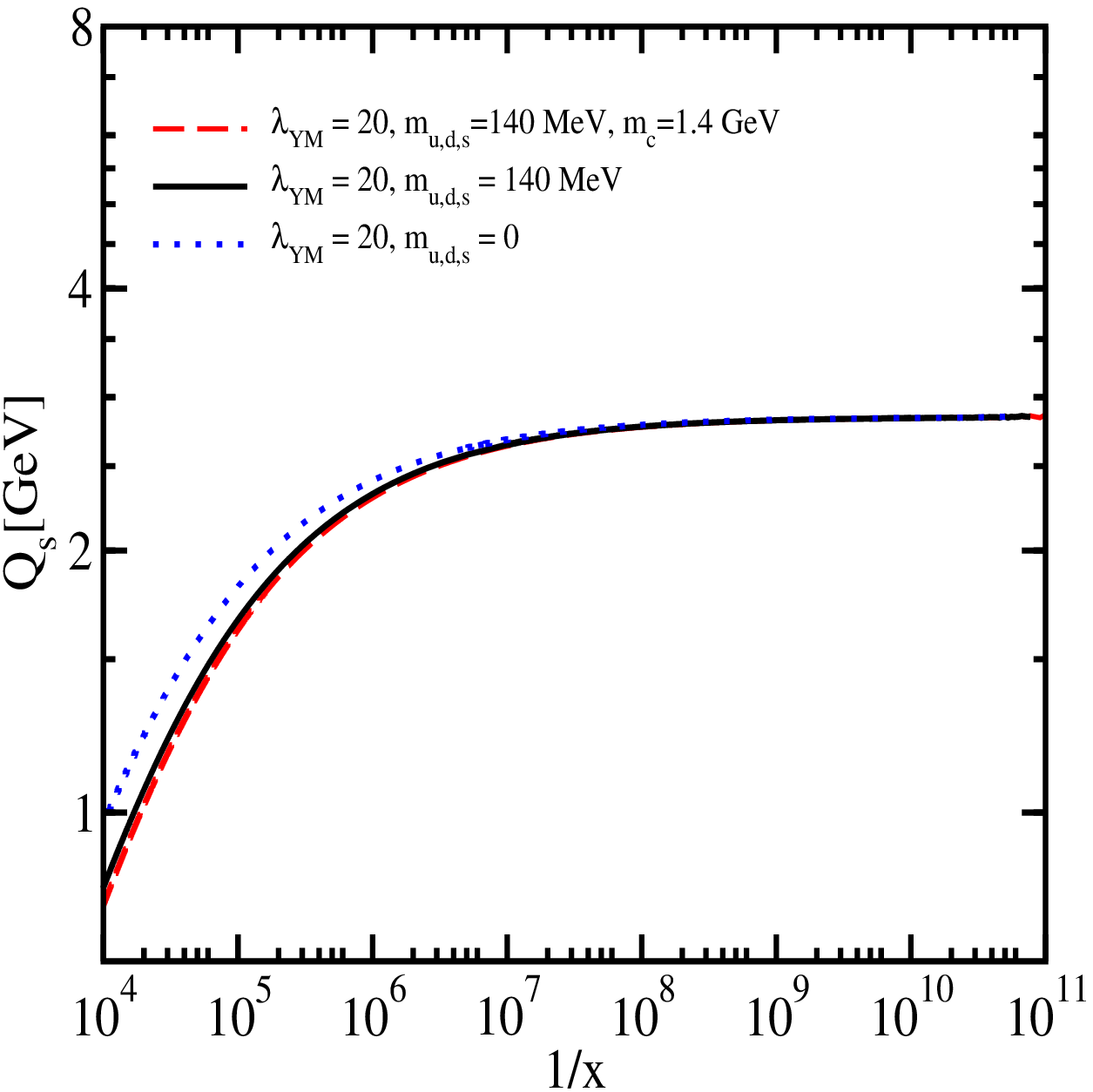} 
   \caption{ The effect of quark mass on saturation scale in the 
     AdS/CFT dipole model. We used the fits given in table III.
     \label{fig22}}
\end{figure}

Let us now plot the $F_2$ structure function given by our fit
  presented above.  In Fig.~\ref{fig1}, we show the description of
the proton structure function $F_2$ obtained from the fit given in
table I for the AdS/CFT and in table II for the GBW dipole model.
Notice that although both models give a good fit of existing data,
they lead to drastically different predictions for the structure
function at smaller $x$ in the region where there is no experimental
data yet. The main prediction of the AdS/CFT color-dipole model is
that at very small $x$ it gives rise to a saturating behavior of the
structure function which becomes independent of $x$. The onset of this
limiting (scaling) behavior moves to a smaller $x$ for larger $Q^2$.
This can be also seen from Fig.~\ref{fig2} where in the left panel we
plot the AdS/CFT dipole cross-section as a function of the dipole
transverse size $r$. It is obvious that AdS/CFT dipole cross-section
profile saturates for $x<10^{-8}$ and will not change further with
$x$.  This is in contrast with the GBW model (and other available
dipole models) where the dipole cross-section rapidly changes as we
move toward smaller $x$, though a certain slowing down of the
$x$-dependence at small-$x$ is observed in \cite{Albacete:2009fh} due
to running coupling effects.  In the right panel of Fig.~\ref{fig2} we
show the saturation scale for both the AdS/CFT and the GBW dipole
models.  It can be seen that the saturation scale in AdS/CFT dipole
model is smaller than the one obtained from the GBW model at very
small $x$.  Moreover, the AdS/CFT model of \cite{ads1} predicts that
the saturation scale saturates.  The saturation scale in the AdS/CFT
dipole model defined via Eq.~(\ref{qads}) is proportional to
$\sqrt{\lambda_{YM}}/\mathcal{M}_0^2$. Therefore, smaller
$\lambda_{YM}$ leads to a smaller saturation scale. One can also
  see from Fig.~\ref{fig22} that the saturation scales for the case of
  massless light flavors only ($m_{u,d,s} = 0$) and also for the case
  of massive light quarks with $m_{u,d,s}=140$~MeV with the charm
  quark included are very similar to the case of three light flavors
  only with $m_{u,d,s}=140$ MeV. This independence of the saturation
  scale of the light quark masses is due to the large saturation scale
  in this model, which cuts off the infrared effects making the
  physics less sensitive to the $u,d,s$ quark masses. The fact that
  the saturation scale in \fig{fig22} is rather large probably
  explains why the charm quark mass does not affect it either. The
curves for the structure function $F_2$ in the massless light quarks
case are not very different visually from the massive quarks case
curves shown in Fig.~\ref{fig1} already: that is why we do not show
the massless case curves there.

\begin{figure}[!t]
  \includegraphics[width=8 cm] {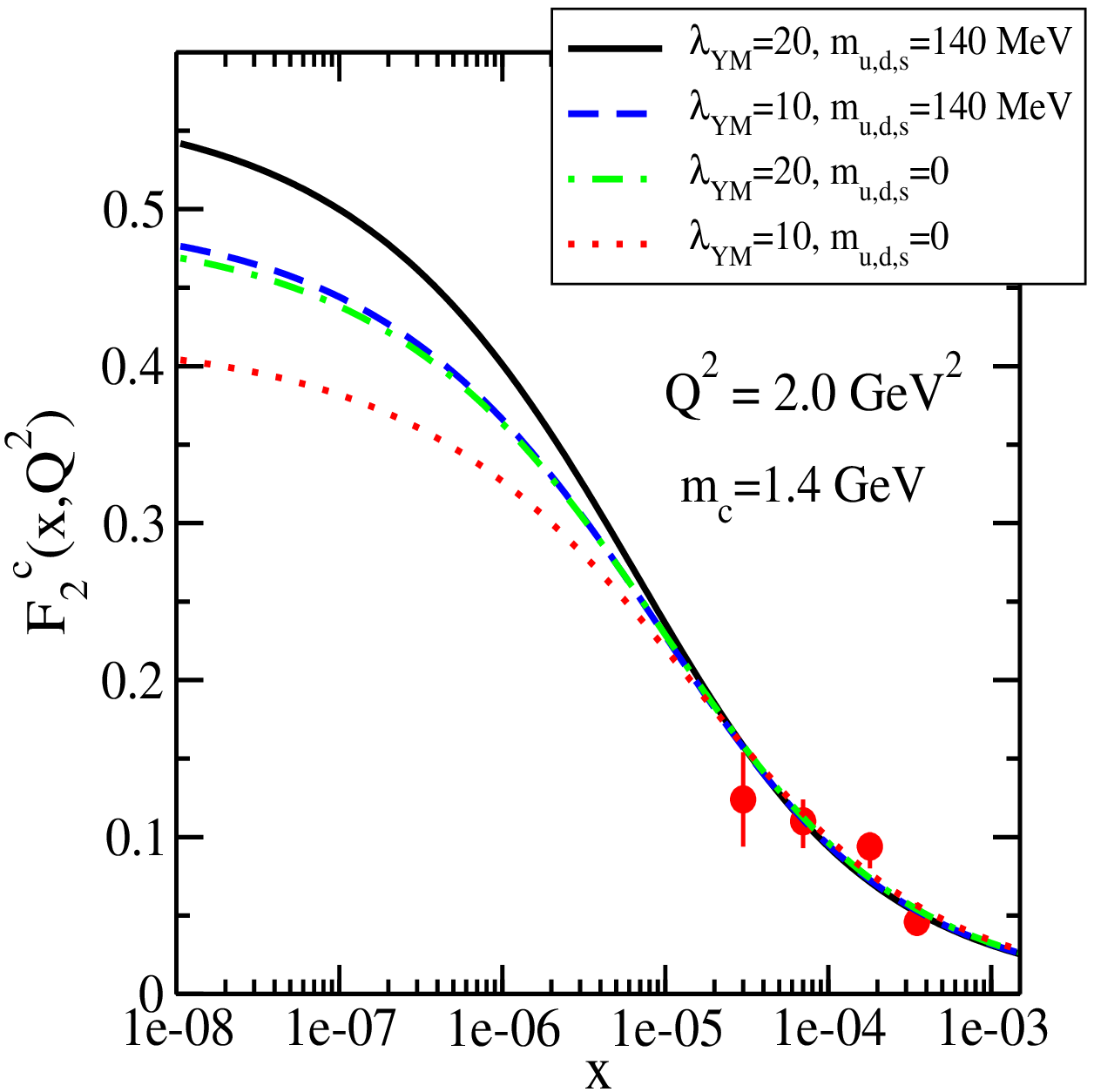}
  \includegraphics[width=8 cm] {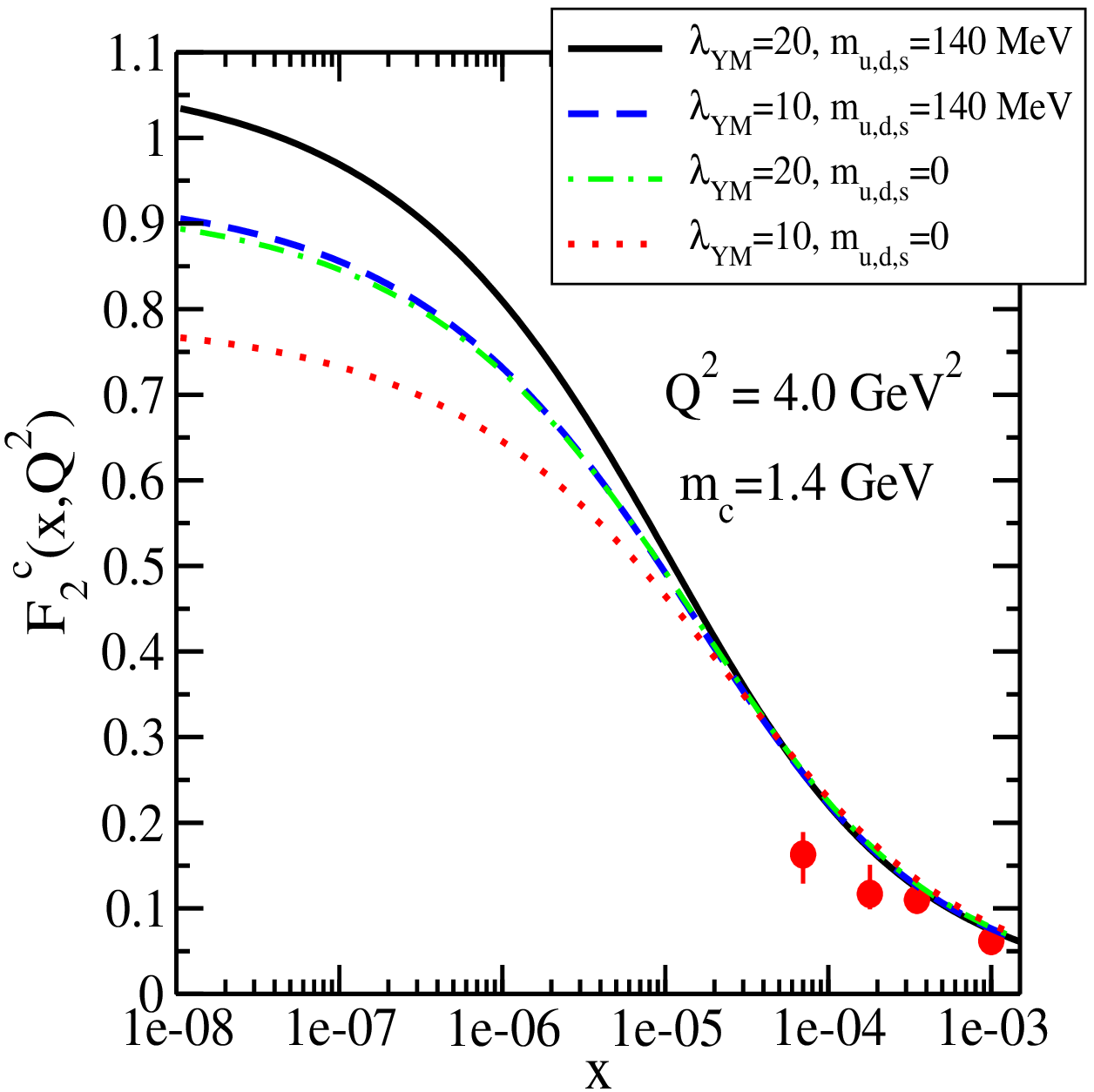}
  \caption{Predictions of AdS/CFT dipole model for the charm structure
    function $F_2^c(x,Q^2)$ plotted as a function of $x$ for various
    values of $Q^2$ shown on the right side of each panel.
    Experimental data are from ZEUS collaboration \cite{zzzd}. The
    curves are generated by the fits given in table III with $m_c=1.4$
    GeV and different values of light quark masses indicated in the
    legend.  }
  \label{fig33}
\end{figure}

In Fig.~\ref{fig33}, we plot the charm structure function
  $F_2^c(x, Q^2)$ given by our AdS/CFT dipole model. Note that we use
  a fit to $F_2$ data (table III) within the range of $x \in[6.2\times
  10^{-7},6\times 10^{-5}]$ and $Q^2/\text{GeV}^2\in [0.045,2.5]$.
  Therefore the experimental data in Fig.~\ref{fig33} are beyond the
  range of our fit. Moreover, the large values of $Q^2$ in \fig{fig33}
  push our AdS-inspired model to the limit of its validity. Hence the
  curves in \fig{fig33} can be thought of as predictions of our model.
  We see that the agreement with data even in this region is rather
  good.

\begin{figure}[!t]
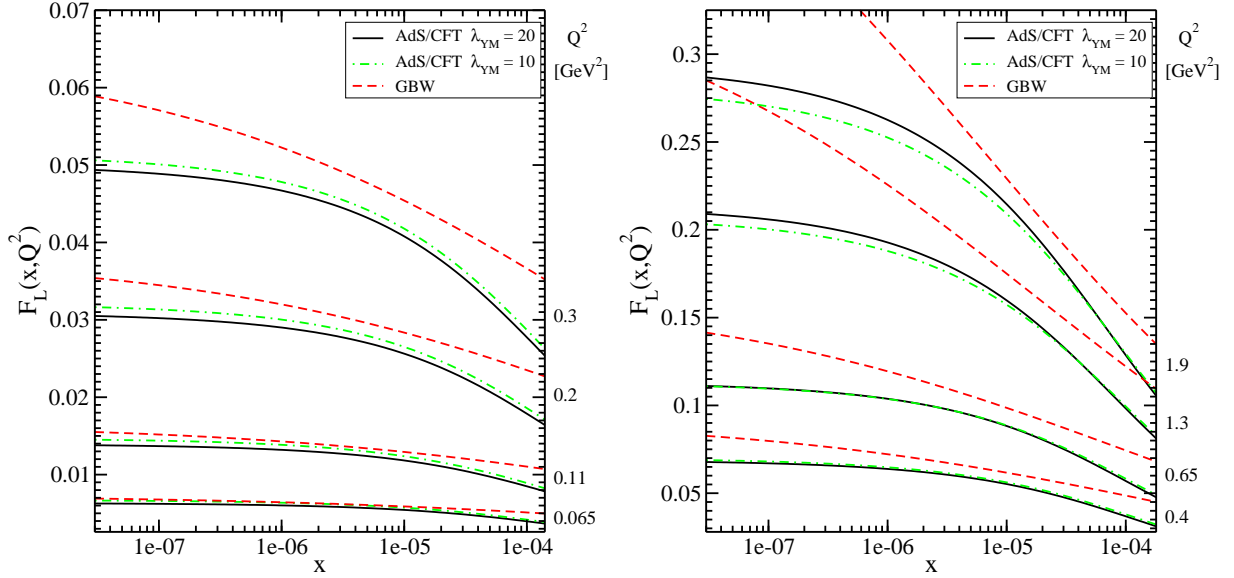

  \includegraphics[width=8 cm] {plot-FL1.eps} 
  \includegraphics[width=8 cm] {plot-FL2.eps}
 \caption{Predictions for the longitudinal structure function $F_L(x,Q^2)$ 
   versus $x$ for various $Q^2$ shown on the right margin of each
   panel. The curves are generated by the same fits as in
   Fig.~\ref{fig1}.}
      \label{fig3}
\end{figure}

 In Fig.~\ref{fig3}, we show the predictions of our AdS-inspired
  model for the longitudinal structure function $F_L(x, Q^2)$
  calculated using \eq{FL}. We use the same fits as employed in
  Fig.~\ref{fig1}. Unfortunately currently there is no data for
  $F_L(x, Q^2)$ at low $Q^2$ and low $x$ where our model is valid. It
  can be seen from Fig.~\ref{fig3} that a precise measurement of $F_L$
  at small $x$ and $Q^2$ can offer a complimentary information which
  may help one discriminate between different DIS models.

\begin{figure}[!t]
  \includegraphics[width=8 cm] {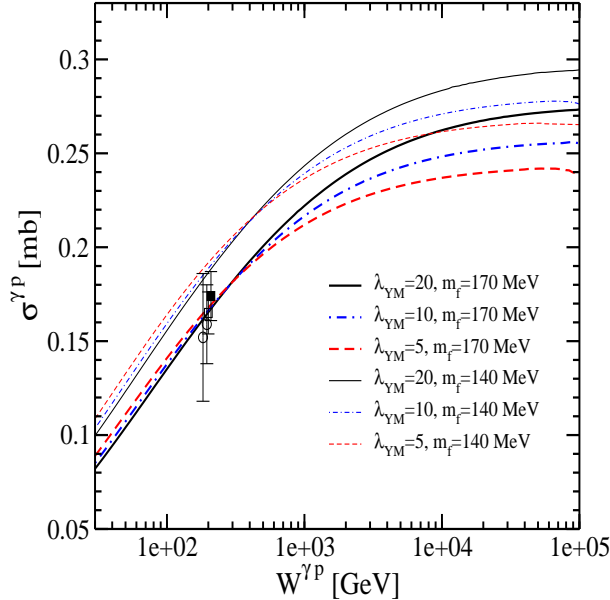} 
 \caption{Predictions of our model for total photoproduction cross-section 
   $\sigma^{\gamma \, p}$ (i.e., $\sigma^{\gamma^*p}$ at $Q^2=0$).
   Experimental data are from H1 \cite{fdata1} and ZEUS \cite{fdata2}
   collaborations. We used the fits given in table V for two different
   values of the light quark mass $m_{f=u,d,s}$ without including
   charm quark. Note that we did not use the photoproduction data to
   generate the fit: the curves shown were not fitted to the data
   points.}
 \label{fig4}
\end{figure}

In Fig.~\ref{fig4}, we show our predictions for the total
photoproduction cross section $\sigma^{\gamma \, p}$. It is calculated
by taking $\sigma^{\gamma^*p}_T$ at $Q^2=0$. (As can be seen from Eqs.
(\ref{gp}) and (\ref{wavL}), $\sigma^{\gamma^*p}_L =0$ at $Q^2=0$ and
does not need to be included.) One can see that using the same
effective quark mass $m_{u,d,s} = 140$~MeV as in the $F_2$ fit our
model slightly overestimates photoproduction data, though mostly
remains withing the error bars of the data points. To show the effect
of light quark mass $m_{u,d,s}$ on photoproduction cross section we
also show the predictions of our model for $m_{u,d,s} = 170$~MeV,
which go directly through the photoproduction data. Indeed using
$m_{u,d,s} = 170$~MeV would lead to larger $\chi^2$ of the $F_2$ fit
presented above. Therefore, the real predictions of our model are for
$m_{u,d,s} = 140$~MeV and slightly miss photoproduction data. (We
checked that including charm quark will not improve the fit either.)
There could be several reasons for this small discrepancy, one of them
being that, after all, the AdS/CFT calculation \cite{ads1} was not
done for QCD, but for a different theory, ${\cal N} = 4$ SYM. For
photoproduction the most important difference between QCD and ${\cal
  N} = 4$ SYM theory is probably the absence of confinement in the
latter. As one can see from Eqs.  (\ref{gp}) and (\ref{wav}), the
integrand of \eq{gp} falls off as $1/r^2$ for $1/Q_s < r < 1/a_f$ and
decays exponentially ($\propto e^{- 2 \, a_f \, r}$) for $r > 1/a_f$.
(The exponential falloff is due to the light-cone wavefunction
(\ref{wav}) which contains the modified Bessel function $K_1$ which
decays exponentially at large values of the argument.) Exponential
decay is essential for convergence of the integral over $r$ in
\eq{gp}. We see that the effective infrared cutoff of the $r$-integral
is $1/a_f$ and the resulting cross section depends logarithmically on
$a_f$. In case of photoproduction (for $Q^2=0$) we have $a_f=m_f$.
Therefore the non-perturbative light quark mass $m_f$ serves as the
only infrared cutoff of the $r$-integral in \eq{gp} in the
photoproduction ($Q^2=0$) limit, as the photoproduction cross section
becomes infinite for $m_f =0$.  Hence in QCD photoproduction cross
section is dominated by non-perturbative effects: this is the basis
for the vector meson dominance models.  While AdS/CFT calculation
\cite{ads1} does indeed contain non-perturbative effects, it is done
for a theory without confinement, allowing for the slight disagreement
between our $m_{u,d,s} = 140$~MeV curves and the data in \fig{fig4}.

Note that we have not included total photoproduction cross-section
data in obtaining the fit given in table V since there are only few
data points at high energies with rather large error bars. The
lower-energy photoproduction data points, while exist \cite{fdata1},
are beyond the limit of applicability of our high-energy model. It has
been already shown that it is very difficult to simultaneously
describe the low energy photoproduction total cross-section and $F_2$
data with a single color dipole amplitude without any extra input
\cite{jeff}.  This is partly due to the fact that the color dipole
approach is not valid at low energy as the coherence length becomes
too short compared to the size of the target proton. One should also
bear in mind that the AdS/CFT color dipole amplitude was derived while
modeling the proton by an ultra-relativistic shock wave, which may not
be a good approximation for lower energy scattering. 


From Figs.~\ref{fig1}, \ref{fig33}, \ref{fig3} and \ref{fig4} it is
again clear that the AdS/CFT color dipole model predicts that at low
virtuality and at very high energy/very small-$x$, the underlying
dipole-target cross section $\sigma_{q \bar q}$ should become
independent of $x$ or $s$, leading to a plateau in the $x$-dependence
for $F_2$, $F_2^c$, $F_L$ and the total photoproduction
cross-section.


\section{Discussion}

In this paper we have demonstrated that the AdS/CFT-inspired
parameterization of the dipole amplitude $N (r,x)$ is consistent with
the existing low-$Q^2$ HERA data for the $F_2$ structure function. The
AdS/CFT parameterization of the dipole amplitude allows to make
distinct predictions for $F_2$ and {\bf $F_L$} structure functions at
values of $x$ below those where the data exist. In particular our
AdS/CFT-inspired model predicts {\em $x$-independence} of the
structure functions $F_2$ and {\bf $F_L$} at very small-$x$. Hence the
predictions of our AdS/CFT parameterization can be tested at the
future colliders, such as LHC and the proposed LHeC.

Indeed to make the above AdS/CFT model fit the data we had to assume
that $c_0$ is about two orders of magnitude smaller than $c_0 \approx
0.83$ predicted by classical AdS/CFT calculations of \cite{ads1}. The
discrepancy between the two results may be due to the difference
between $\mathcal{N}=4$ SYM theory and QCD. One should also remember
that the calculation of \cite{ads1} was purely classical (extremizing
the string profile in the classical gravity background), and quantum
corrections of the order of $1/\sqrt{\lambda_{YM}}$ may be important
for the description of the data. Further research is needed to quantify
these issues.

A more general question about the applicability of strong-coupling
based methods like AdS/CFT to the description of DIS data also has to
be asked. Indeed if the strong coupling constant always runs with
$Q^2$, then at low-$Q^2$ considered above the coupling should be large
justifying the use of non-perturbative approaches. At the same time at
low-$Q^2$ but with $x$ small enough for $Q_s$ to be large, it is
likely that the coupling runs with the saturation scale $Q_s$. As the
proton's saturation scale for the range of small $x$ considered above
varies in the interval of $1 \div 3$~GeV for most models
\cite{gbw,Albacete:2007sm,Albacete:2009fh}, one could then argue that
the problem is perturbative and strongly-coupled methods are not
needed to describe the DIS data. Indeed purely perturbative CGC
approaches are rather successful in describing the DIS data (see
\cite{Albacete:2009fh} for the most comprehensive and rigorous CGC
calculation to date). However, it is likely that the story is more
complicated: as one can see in the explicit running coupling
calculations \cite{rc} for the BK and JIMWLK equations, the strong
coupling runs with the size of the dipoles, which indeed varies from
non-perturbative to perturbative distance scales. Hence even at large
$Q_s$ the non-perturbative contribution to $F_2$ may be
non-negligible, though it does tend to be suppressed as $Q_s$ grows
very large. Our work above could be viewed as an effort to estimate
the shape of the contribution of the non-perturbative physics to the
$F_2$ structure function. We find it rather interesting that, modulo
the above-mentioned open questions, the non-perturbative
AdS/CFT-inspired physics can be made largely consistent with the $F_2$
data at small-$Q^2$.


\begin{acknowledgments}
  
  Yu.K. would like to thank Boris Kopeliovich, Irina Potashnikova, and
  the Theoretical Physics group at the Universidad T\'ecnica Federico
  Santa Mar\'\i a in Valpara\'\i so, Chile for their hospitality and
  support during the initial stages of this work. Yu.K. is also
  grateful to Genya Levin for informative discussions. The work of
  Yu.K. is sponsored in part by the U.S. Department of Energy under
  Grant No.  DE-FG02-05ER41377.
  
  The work of Z.L. is supported by PBCT (Chile) project ACT-028.  
   
  A.H.R. would like to thank Boris Kopeliovich for useful discussions.
  A.H.R. is very grateful to the hospitality of Andreas Sch\"afer
  group at Universit\"at Regensburg where this work was finalized, and
  acknowledges the financial support from Conicyt Programa
  Bicentenario PSD-91-2006, Fondecyt grants 1090312 (Chile) and the
  Alexander von Humboldt foundation (Germany).
  
  The authors are grateful to Javier Albacete for a useful suggestion.

\end{acknowledgments}  



\begin{thebibliography}{99}


\bibitem{Gribov:1984tu}
L.~V. Gribov, E.~M. Levin, and M.~G. Ryskin, 
   {\em Phys. Rept.} {\bf 100} (1983) 1; A.~H. Mueller and J.-w. Qiu,  {\em Nucl. Phys.} {\bf B268} (1986) 427; L.~D. McLerran and R.~Venugopalan,  {\em Phys. Rev.} {\bf D50} (1994) 2225; L.~D. McLerran and R.~Venugopalan,  {\em Phys. Rev.} {\bf D49}
  (1994) 3352; L.~D. McLerran and R.~Venugopalan,  {\em Phys. Rev.} {\bf D49} (1994)
  2233; Y.~V. Kovchegov, {\em
  Phys. Rev.} {\bf D54} (1996) 5463; Y.~V. Kovchegov,   {\em Phys. Rev.} {\bf D55} (1997) 5445; 
J.~Jalilian-Marian, A.~Kovner, L.~D. McLerran, and H.~Weigert, {\em Phys. Rev.} {\bf D55}
  (1997) 5414.  



\bibitem{Kovchegov:1999yj}
Y.~V. Kovchegov,   {\em Phys. Rev.} {\bf D60} (1999) 034008; Y.~V. Kovchegov,   {\em
  Phys. Rev.} {\bf D61} (2000) 074018; I.~Balitsky,  {\em Nucl.
  Phys.} {\bf B463} (1996) 99; I.~Balitsky, hep-ph/9706411; I.~Balitsky,  {\em Phys.
  Rev.} {\bf D60} (1999) 014020. 

\bibitem{Jalilian-Marian:1997jx}
J.~Jalilian-Marian, A.~Kovner, A.~Leonidov, and H.~Weigert,  {\em Nucl. Phys.} {\bf
  B504} (1997) 415; J.~Jalilian-Marian, A.~Kovner, A.~Leonidov, and H.~Weigert, 
  {\em Phys. Rev.} {\bf D59} (1998) 014014; J.~Jalilian-Marian, A.~Kovner, and H.~Weigert,   {\em Phys. Rev.} {\bf D59} (1998) 014015; J.~Jalilian-Marian, A.~Kovner, A.~Leonidov, and H.~Weigert, 
  {\em Phys. Rev.} {\bf D59} (1999) 034007; A.~Kovner, J.~G. Milhano, and H.~Weigert, {\em Phys. Rev.} {\bf
  D62} (2000) 114005; H.~Weigert,  {\em Nucl. Phys.} {\bf A703}
  (2002) 823; E.~Iancu, A.~Leonidov, and L.~D. McLerran,  {\em Nucl. Phys.} {\bf A692} (2001)
  583; 
E.~Ferreiro, E.~Iancu, A.~Leonidov, and L.~McLerran,  {\em Nucl. Phys.} {\bf A703}
  (2002) 489. 

\bibitem{Iancu:2003xm}
E.~Iancu and R.~Venugopalan, hep-ph/0303204; H.~Weigert,   {\em Prog. Part. Nucl. Phys.} {\bf 55} (2005) 461; J.~Jalilian-Marian and Y.~V. Kovchegov, {\em Prog. Part. Nucl. Phys.} {\bf 56} (2006)
  104. 

\bibitem{Kuraev:1977fs}
E.~A. Kuraev, L.~N. Lipatov, and V.~S. Fadin, {\em Sov. Phys. JETP} {\bf 45}
  (1977) 199; Y.~Y. Balitsky and L.~N. Lipatov {\em Sov. J. Nucl. Phys.} {\bf 28} (1978) 822.


\bibitem{Iancu:2002tr}
E.~Iancu, K.~Itakura, and L.~McLerran,  {\em Nucl. Phys.} {\bf A708} (2002) 327. 


\bibitem{Albacete:2004gw}
J.~L. Albacete, N.~Armesto, J.~G. Milhano, C.~A. Salgado, and U.~A. Wiedemann, {\em Phys. Rev.} {\bf D71} (2005) 014003. 


\bibitem{Fadin:1998py}
V.~S. Fadin and L.~N. Lipatov, {\em Phys. Lett.} {\bf B429} (1998) 127; M.~Ciafaloni and G.~Camici,  {\em Phys. Lett.} {\bf B430} (1998) 349; 
I.~Balitsky and G.~A. Chirilli,  {\em Phys. Rev.} {\bf D77} (2008) 014019; 
D.~A. Ross,  {\em Phys. Lett.} {\bf B431} (1998) 161; 
Y.~V. Kovchegov and A.~H. Mueller,  {\em Phys. Lett.} {\bf B439} (1998) 428. 


\bibitem{Albacete:2007sm}
J.~L. Albacete, {\em Phys. Rev.  Lett.} {\bf 99} (2007) 262301;
M.~Ciafaloni and G.~Camici, {\em Phys. Lett.} {\bf B430} (1998) 349;
I.~Balitsky and G.~A. Chirilli, {\em Phys. Rev.} {\bf D77} (2008)
014019; D.~A. Ross, {\em Phys. Lett.} {\bf B431} (1998) 161;
Y.~V. Kovchegov and A.~H. Mueller, {\em Phys. Lett.} {\bf B439} (1998)
428; M.~Ciafaloni, D.~Colferai, and G.~P. Salam, {\em Phys. Rev.} {\bf
D60} (1999) 114036. 

\bibitem{rc}
I.~I. Balitsky, {\em Phys. Rev. D} {\bf 75} (2007)
014001; Y.~Kovchegov and H.~Weigert, {\em Nucl. Phys. {\bf A}} {\bf
784} (2007) 188; E.~Gardi, J.~Kuokkanen, K.~Rummukainen, and
H.~Weigert, {\em Nucl. Phys.} {\bf A784} (2007) 282;  
J.~L. Albacete and Y.~V. Kovchegov,  {\em Phys. Rev.} {\bf D75} (2007)
  125021. 

\bibitem{Albacete:2009fh}
  J.~L.~Albacete, N.~Armesto, J.~G.~Milhano and C.~A.~Salgado,
  arXiv:0902.1112.


\bibitem{Maldacena:1997re}
J.~M. Maldacena, {\em Adv. Theor. Math. Phys.} {\bf 2} (1998) 231;
S.~S. Gubser, I.~R. Klebanov, and A.~M. Polyakov, {\em Phys. Lett.} 
{\bf B428} (1998) 105; E.~Witten, {\em Adv. Theor. Math.  Phys.} {\bf 2} (1998) 253;
O.~Aharony, S.~S. Gubser, J.~M. Maldacena, H.~Ooguri, and Y.~Oz, {\em
Phys. Rept.} {\bf 323} (2000) 183.

\bibitem{Janik:1999zk}
  R.~A.~Janik and R.~B.~Peschanski,
  Nucl.\ Phys.\  B {\bf 565}, 193 (2000)
  [arXiv:hep-th/9907177].



\bibitem{Polchinski:2001tt}
  J.~Polchinski and M.~J.~Strassler,
  Phys.\ Rev.\ Lett.\  {\bf 88}, 031601 (2002)
  [arXiv:hep-th/0109174];
  R.~C.~Brower, J.~Polchinski, M.~J.~Strassler and C.~I.~Tan,
  JHEP {\bf 0712}, 005 (2007)
  [arXiv:hep-th/0603115].


\bibitem{Hatta:2007cs}
  Y.~Hatta, E.~Iancu and A.~H.~Mueller,
  JHEP {\bf 0801}, 063 (2008)
  [arXiv:0710.5297];
  L.~Cornalba and M.~S.~Costa,
  Phys.\ Rev.\  D {\bf 78}, 096010 (2008)
  [arXiv:0804.1562];
  P.~G.~O.~Freund and H.~Nastase,
  arXiv:0809.1277;
  E.~Levin, J.~Miller, B.~Z.~Kopeliovich and I.~Schmidt,
  JHEP {\bf 0902}, 048 (2009)
  [arXiv:0811.3586].


\bibitem{Donnachie:1998gm}
  A.~Donnachie and P.~V.~Landshoff,
  Phys.\ Lett.\  B {\bf 437}, 408 (1998)
  [arXiv:hep-ph/9806344].


\bibitem{ads1}
 J. L. Albacete, Y. V. Kovchegov and A. Taliotis,  {\em JHEP } {\bf 0807} (2008) 074. 
 

\bibitem{Nikolaev:1990ja}
  N.~N.~Nikolaev and B.~G.~Zakharov,
  Z.\ Phys.\  C {\bf 49}, 607 (1991).

\bibitem{gbw} K. Golec-Biernat and M. W\"usthoff, {\em Phys. Rev.}
  {\bf D59} (1999) 014017; {\bf D60} (1999) 114023.

\bibitem{Kovchegov:1999kx}
  Y.~V.~Kovchegov and L.~D.~McLerran,
  Phys.\ Rev.\  D {\bf 60}, 054025 (1999)
  [Erratum-ibid.\  D {\bf 62}, 019901 (2000)]
  [arXiv:hep-ph/9903246].

\bibitem{Lublinsky:2001yi}
  M.~Lublinsky, E.~Gotsman, E.~Levin and U.~Maor,
  Nucl.\ Phys.\  A {\bf 696}, 851 (2001)
  [arXiv:hep-ph/0102321];
  E.~Levin and M.~Lublinsky,
  Nucl.\ Phys.\  A {\bf 696}, 833 (2001)
  [arXiv:hep-ph/0104108].


\bibitem{gbw-g}
 L. Motyka, K. Golec-Biernat and G. Watt, arXiv:0809.4191;  B. Z. Kopeliovich, A. H. Rezaeian, arXiv:0811.2024;
 B. Z. Kopeliovich, E. Levin, A. H. Rezaeian and I. Schmidt, {\em Phys. Lett.} {\bf B675} (2009) 190 [arXiv:0902.4287].

\bibitem{Maldacena:1998im}
  J.~M.~Maldacena,
  Phys.\ Rev.\ Lett.\  {\bf 80}, 4859 (1998)
  [arXiv:hep-th/9803002].


\bibitem{di1}
 G. Watt and H. Kowalski, {\em Phys. Rev.} {\bf D78} (2008) 014016.
\bibitem{di2}
 H. Kowalski, L. Motyka and G. Watt, {\em Phys. Rev.} {\bf D74} (2006) 074016.
\bibitem{di3}
 H. Kowalski and D. Teaney, {\em Phys. Rev.} {\bf D68} (2003) 114005.


\bibitem{Stasto:2000er}
  A.~M.~Stasto, K.~J.~Golec-Biernat and J.~Kwiecinski,
  Phys.\ Rev.\ Lett.\  {\bf 86} (2001) 596
  [arXiv:hep-ph/0007192].

\bibitem{bk-g}
D. Boer, A. Utermann and E. Wessels, {\em Phys. Rev.}  {\bf D75} (2007) 094022.

\bibitem{di4}
 E. Iancu, K. Itakura and S. Munier, {\em Phys. Lett.} {\bf B590} (2004) 199.

\bibitem{zu1}
  J.~Breitweg {\it et al.}  [ZEUS Collaboration],
 {\em  Eur. Phys. J.}  {\bf C7} (1999) 609 [arXiv:hep-ex/9809005].



\bibitem{zu2}
  M.~Derrick {\it et al.}  [ZEUS Collaboration],
 {\em  Z.\ Phys.\  } {\bf C69} (1996) 607 
  [arXiv:hep-ex/9510009].


\bibitem{zu3}
  S.~Chekanov {\it et al.}  [ZEUS Collaboration],
  {\em Eur.\ Phys.\ J.\  } {\bf C21} (2001) 443 
  [arXiv:hep-ex/0105090].



\bibitem{zu4}
  J.~Breitweg {\it et al.}  [ZEUS Collaboration],
  {\em Phys.\ Lett.\ }  {\bf B487} (2000) 53 
  [arXiv:hep-ex/0005018].



\bibitem{di5}
J. R. Forshaw and G. Shaw, {\em JHEP} {\bf 0412} (2004) 052. 

\bibitem{lqcd}
D. N. Triantafyllopoulos, {\em Nucl. Phys.} {\bf B648} (2003) 293.

\bibitem{lbk}
A. H. Mueller and D.N. Triantafyllopoulos, {\em Nucl. Phys.} {\bf B640} (2002) 331 ;
D. N. Triantafyllopoulos, {\em Nucl. Phys.} {\bf B648}  (2003) 293.


\bibitem{Gubser:2006qh}
  S.~S.~Gubser,
  Phys.\ Rev.\   {\bf D76} (2007) 126003
  [arXiv:hep-th/0611272].

\bibitem{zzzd}
ZEUS Collaboration, {\em Phys. Rev.} {\bf D69} (2004) 012004.

\bibitem{fdata1}
H1-collaboration, {\em Phys. Lett.} {\bf B299} (1993) 374; {\em Z. Phys.} {\bf C69} (1995) 27.

\bibitem{fdata2}
ZEUS Collaboration, {\em Nucl. Phys.} {\bf B627} (2002) 3. 

\bibitem{jeff}
J. R. Forshaw, G. Kerley and G. Shaw, Phys. Rev. {\bf D60} (1999) 074012. 

\end{thebibliography}
\end{document}